# How Charge Carrier Exchange between Absorber and Contact influences Time Constants in the Frequency Domain Response of Perovskite Solar Cells

Sandheep Ravishankar,[1*] Zhifa Liu,[1] Yueming Wang,[1] Thomas Kirchartz[1,2] and Uwe Rau[1]

[1]IEK-5 Photovoltaik, Forschungszentrum Jülich, 52425 Jülich, Germany
[2]Faculty of Engineering and CENIDE, University of Duisburg-Essen, Carl-Benz-Str. 199, 47057 Duisburg, Germany

*author for correspondence, email: s.ravi.shankar@fz-juelich.de

**Abstract**

A model is derived for the frequency- and time-domain opto-electronic response of perovskite solar cells (PSCs) that emphasizes the role of charge carrier exchange, .i.e. extraction and injection, from (to) the perovskite through the transport layer to (from) the collecting electrode. This process is described by a charge carrier exchange velocity that depends on the mobility and electric field inside the transport layer. The losses implied by this process are modelled in an equivalent circuit model in the form of a voltage-dependent transport layer resistance. The analysis of the model predicts that the voltage dependence of the measured time constants allows discriminating situations where the transport layer properties dominate the experimental response. Application of this method to experimental impedance spectroscopy data identifies charge extraction velocities between 1-100 cm/s at 1 sun open-circuit conditions for *p-i-n* PSCs with *PTAA* as the hole transport layer, that corresponds to transport layer mobilities between $10^{-4}$-$3\times10^{-3}$ $cm^2V^{-1}s^{-1}$. The model paves the way for accurate estimation of photocurrent and fill factor losses in PSCs caused by the low mobilities in the transport layers, using small perturbation measurements in the time and frequency domain.

1. **Introduction**

The analysis of different fundamental working mechanisms in perovskite solar cells (PSCs) is largely carried out using a suite of time domain and frequency domain opto-electronic techniques. These techniques generally involve the measurement of the response of the solar cell or stack to a perturbation of the photon flux $\phi$, the external voltage $V_{ext}$ or the current density $j$. In the time domain, these techniques include transient photovoltage (TPV), transient photocurrent (TPC) and transient photoluminescence (tr-PL). While TPV and TPC measure the external voltage and current density respectively to a small perturbation of light intensity, tr-PL measures the emitted photon flux to a large perturbation of light intensity. The analysis of these measurements typically involves the fitting of the rise or decay of the measured quantity using a mono or bi-exponential fit,[1-4] subsequently assigning the calculated time constant to a physical process such as recombination or transport. However, as shown using drift-diffusion simulations, such an analysis is almost always an over-simplification of a situation where several mechanisms can co-exist, superimpose and vary in relative importance as a function of illumination and bias conditions, structure (device or stack) and material properties.[5,6] For example, a comparison of lifetimes obtained from tr-PL (usually measured on stacks) and TPV (measured on devices) show that the TPV lifetimes are generally a few orders higher due to measurement of the geometric capacitance-dominated region of the apparent lifetime.[7,8] In this regard, in the case of TPV and tr-PL, the definition of a differential decay time at each point of the decay in addition to conversion of the time axis to an open-circuit voltage ($V_{oc}$) axis allows easier discrimination of physical mechanisms, facilitating comparison of different device structures and also between the two methods.

In the case of the frequency domain, the corresponding counterparts for TPV and TPC are intensity-modulated photovoltage spectroscopy (IMVS) and intensity-modulated photocurrent spectroscopy (IMPS), while impedance spectroscopy (IS) measures the modulated current density upon the application of a modulated external voltage. Recent work using these methods has focussed on combining the information contained within each of them to obtain a unified interpretation, in the form of a single equivalent circuit [9,10] or through the application of the diffusion-recombination model.[11,12] While equivalent circuits allow easier modelling and analysis of overlapping physical mechanisms compared to the time domain methods, it is still unclear which equivalent circuit reliably models a wide range of PSCs.

However, the biggest concern in the interpretation of both time domain and frequency domain spectra is the non-accounting of non-linear transport layer effects. While the perovskite is known to be an efficient transporter of charge due to its large electron and hole mobilities (mobility $>1$ cm$^2$V$^{-1}$s$^{-1}$),[13] the transport layers are much less efficient in comparison. Furthermore, due to their low thickness, often low permittivity and typically also low conductivity, they do not behave like a simple series resistance that could easily be implemented into an equivalent circuit. This is especially the case for *p-i-n* PSCs, which employ thin, organic layers such as poly(triaryl amine) (*PTAA*), [6,6]-phenyl-$C_{61}$ butyric acid methyl ester (*PCBM*), $C_{60}$-fused N-methylpyrrolidine-m-$C_{12}$-phenyl (*CMC*) and indene-$C_{60}$ bisadduct (*ICBA*),[14] with comparatively lower mobilities between $10^{-5} - 10^{-2}$ cm$^2$V$^{-1}$s$^{-1}$.[15-17] This means that the effective response in a measurement can contain contributions or even be dominated by the response of the transport layers, as we have shown recently for different capacitance-based methods used to investigate PSCs.[18-20] In such cases, the measured response and hence the calculated parameters are wholly and wrongly attributed to that of the perovskite layer. In addition, recent works have shed light on the existence of an internal series resistance in any illuminated solar cell in general.[21,22] This series resistance is a function of the external voltage and models the translation of the internal voltage $V_{\text{int}}$ to the voltage $V_{\text{ext}}$ at the electrodes, that is required to drive the current density from the bulk to the outer circuit.[23] In the case of PSCs, this resistance is the chemical potential drop in the transport layer (termed the transport layer resistance) and is a function of the mobility of the transport layer and the electric field inside it.[18] Therefore, a new model that accounts for this transport layer resistance and decouples the response of charge extraction and transport in the transport layers from that of recombination in the perovskite layer is required, in order to model the time and frequency domain spectra of PSCs more effectively.

In this work, we solve this problem by the development of a model that is applicable to small perturbation measurements of any solar cell that consists of a high mobility absorber layer in series with charge transport layers of variable mobility. In addition to encompassing the fundamental mechanisms of charge generation, recombination and capacitive discharge, this model explicitly accounts for the difference in internal and external voltage and the associated charge carrier transfer between the bulk and the electrodes, in the form of an injection/extraction velocity $S$ that depends on the quality of the charge transport layers. The model predicts that the voltage dependence of the measured time constants contains additional information that can be exploited to discriminate between the overlapping mechanisms of recombination in the perovskite layer and charge extraction and transport in the transport layers. Analysis of experimental IS measurements of PSCs with different transport layers using this method identifies that the high frequency resistance of *PTAA*-based PSCs at open-circuit voltages at the 1 sun value is dominated by the transport layer resistance. This yields charge carrier exchange velocities between 1-100 cm/s at 1 sun open-circuit conditions, which corresponds to transport layer mobilities between $10^{-4}$-$3\times10^{-3}$ cm$^2$V$^{-1}$s$^{-1}$ for the *PTAA*-based PSCs.

## 2. Experimental spectra

Impedance Spectroscopy (IS) measurements were performed on a range of *p-i-n* PSCs at open-circuit conditions. These PSCs showed efficiencies between 15-20% and consist of different compositions of the perovskite layer and different electron and hole transport layers (ETL and HTL respectively), summarised in table S1 in the supporting information (SI) (see section A1 for experimental details and figure S1 in the SI for current-voltage curves). These include *PTAA* and self-assembled monolayers (*SAM*s) – 2PACz and Me-4PACz for the HTL, and *PCBM/BCP* (Bathocuproine), *PCBM:CMC:ICBA/BCP* and *$C_{60}$/BCP* for the ETL. The Nyquist (real vs imaginary part of the transfer function) plots from these measurements generally yield two arcs (shown in figure S2 in the SI), one at high frequencies above $10^4$ Hz and one at low frequencies in the region of a few hundred Hz to mHz. At large open-circuit voltages, a third arc at low frequencies is observed. For the case of frequency domain measurements, the PSC response above $\cong 10^3$ Hz is considered to be purely electronic in nature.[24] However, conductivity measurements on perovskite pellets and films have calculated ionic transport frequencies (and hence ion-dominated response) of $\cong 10^5$ Hz and below.[25,26] While the discrepancy in the values between frequency domain measurements and conductivity measurements is not well understood, we assume that frequencies of $10^4$ Hz and above (i.e. high frequency arc in the IS spectrum) are dominated by purely electronic phenomena.

Figure 1(a) shows the measured high frequency time constant and its evolution versus open-circuit voltage. At large open-circuit voltages close to the 1 sun open-circuit voltage, a constant value of the time constant is observed, between $2 \times 10^{-7} - 2 \times 10^{-6}$ s. At lower open-circuit voltages, this plateau transitions into an exponentially increasing time constant with reducing open-circuit voltage. For comparison, the measured time constants of a PSC from a tr-PL measurement of a high performance MAPI PSC is shown,[27] which shows a similar exponential increase at lower open-circuit voltages, followed by a plateau and then an exponential decrease with increasing open-circuit voltage, forming an 'S' shape. The IS time constants follow a similar 'S' shape but do not show the exponential drop in the time constant due to the inability to reach such large values of open-circuit voltage compared to the tr-PL measurement. The corresponding resistances and capacitances that form the IS time constants are shown in figures 1(b) and 1(c) respectively. The resistance decreases exponentially followed by either an increased slope or saturation at large open-circuit voltages, while the capacitance shows a saturation at lower open-circuit voltages followed by an exponential increase at larger open-circuit voltages. We note that the resistances change over several orders of magnitude compared to the capacitances, which means the evolution of the time constant is dominated by the resistances.

The plateau region in the time constants is usually interpreted as originating from a first-order recombination process coupled with the chemical capacitance (that leads to a voltage-independent lifetime), associated to either non-radiative recombination in the bulk ($\tau_{\text{SRH}}$), radiative recombination in low injection conditions with bulk doping density $N_\text{D}$ ($\tau = 1/B_{\text{rad}}N_\text{D}$) or surface recombination.[5] The exponential increase in the time constant at lower open-circuit voltages has been found to originate from the coupling of the discharge of the geometric capacitance to the recombination.[5,7,8,28] The influence of a more general dielectric capacitance on the measured lifetimes has been discussed in ref. [29]. However, these interpretations are not sufficient to explain the data as the IS measurement is also sensitive to the transport layer resistances due to the modulated current density flowing in or out of the cell, and is hence strongly affected by them. This logic is given further credence by the saturation of the resistances in figure 1(b), which is not typical of any recombination mechanism. Therefore, there is a fundamental need for an advanced model of the PSC that incorporates the direct and indirect effects of the transport layers on its transient response, that is derived in the next section.

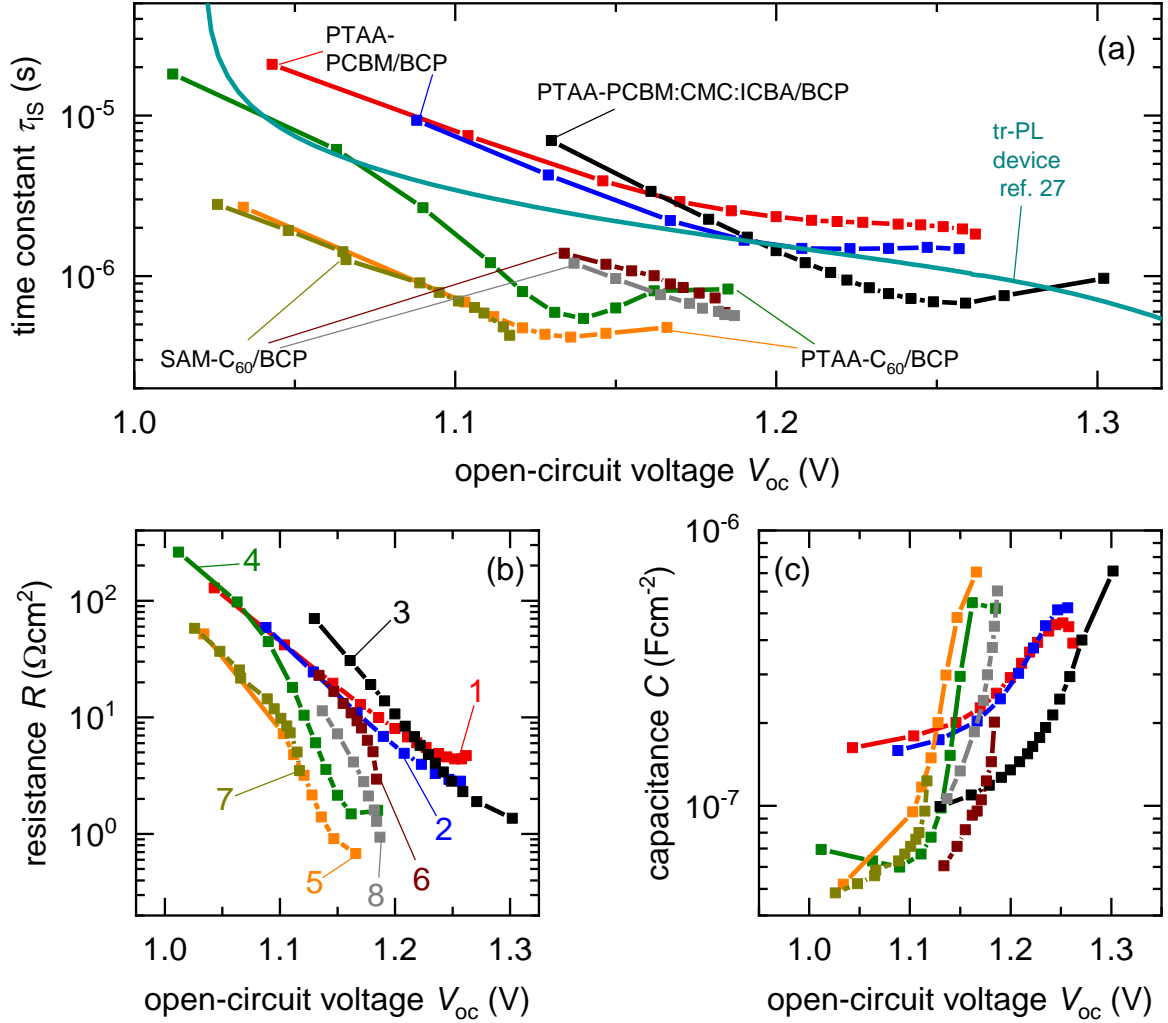

**Figure 1** Measured (a) high frequency time constant and corresponding (b) resistance and (c) capacitance from IS measurements at different open-circuit voltages, on different *p-i-n* PSCs (see table S1 in the SI for summary of sample data and figure S1 in the SI for current-voltage curves). Also shown in (a) using a line with no symbols is the differential lifetime obtained from a tr-PL measurement of a *MAPI* PSC, reproduced with permission from ref.[27]. The labels in (a) indicate the HTL and ETL layers of the cell in the format 'HTL-ETL'. The labels in (b) show the numbers corresponding to each sample (see table S1 in the SI) which are used as reference.

### 3. Theory

#### A. Charge Carrier Exchange between absorber and contact

We begin by assuming a thin-film PSC with an intrinsic perovskite absorber of thickness $d$ in contact with an electron transport layer (ETL) at $x = 0$ of thickness $d_{\text{TL}}$ and a perfectly blocking contact for electrons at $x = d$. We further assume that the electric field inside the perovskite layer is negligible due to shielding by the ions and that the diffusion length is much larger than the thickness of the perovskite layer (a valid assumption for thin-film PSCs[13]), implying a constant and equal density of electrons and holes within it, given by

$$np = n^2 = n_i^2 \exp\left(\frac{qV_{\text{int}}}{k_B T}\right) \rightarrow n = n_i \exp\left(\frac{qV_{\text{int}}}{2k_B T}\right), \tag{1}$$

where $n_i$ is the intrinsic carrier concentration and $V_{\text{int}} = (E_{\text{Fn}} - E_{\text{Fp}})/q$ represents the steady-state splitting of the quasi-Fermi levels $E_{\text{Fn}}, E_{\text{Fp}}$ of electrons and holes inside the absorber layer. Assigning the split of the quasi-Fermi levels to an internal voltage $V_{\text{int}}$ is helpful but we have to bear in mind that this internal 'voltage' is not a voltage in the sense of an *electrostatic* potential, rather it expresses the sum of the excess *chemical* potentials of electrons and holes. The steady-state current $j$ through the device, in the traditional formalism, is given by

$$j = j_{\text{sc}} - j_0 \left( \exp\left( \frac{qV_{\text{elec}}}{n_{\text{id}} k_B T} \right) - 1 \right), \tag{2}$$

where $j_{\text{sc}}$ is the short-circuit current density, $j_0$ is the reverse saturation current density, $n_{\text{id}}$ is the ideality factor and $V_{\text{elec}}$ is the steady-state quasi-Fermi level of the electrons at the cathode minus the quasi-Fermi level of the holes at the anode. Equation 2 connects the current density $j$ to the applied external voltage by assuming that the splitting of the quasi-Fermi levels equals the voltage $V_{\text{elec}}$, i.e., $V_{\text{int}} = V_{\text{elec}}$. Equation (2) is the basis of the superposition principle – adding the photogenerated current to the dark recombination current to produce the current-voltage curve under illumination.

However, while $V_{\text{int}} = V_{\text{elec}}$ might be a good approximation at open circuit, it is never a good approximation for $V_{\text{elec}}$ significantly smaller than $V_{\text{oc}}$. The need to drive current out of a device leads to deviations from the superposition principle that become visible in the form of variations in recombination current under dark and illumination conditions,[22,30] with large luminescence observed at short-circuit conditions for silicon,[31] Cu(In,Ga)Se$_2$ (CIGS)[23], organic solar cells[32] and perovskite solar cells.[22,33] These findings imply large quasi-Fermi level splitting inside the absorber layer under illumination followed by a chemical potential drop to $V_{\text{elec}}$, which can occur either inside the space-charge region (as in silicon or CIGS solar cells) or within the transport layers (as in perovskite solar cells).

The connection between the concentration $n$ of charge carriers inside a photovoltaic absorber and the external voltage $V_{\text{elec}}$ was described in Ref.[34] with the help of an exchange coefficient. Using Eqs.(4) and (5) from Ref.[23] for the calculation of the exchange current density $j_{\text{exc}}$, we find

$$j_{\text{exc}} = qS_{\text{exc}} \left[ n - n_i \exp\left( \frac{qV_{\text{elec}}}{k_B T} \right) \right], \tag{3}$$

where $S_{\text{exc}}$ denotes an exchange velocity for the charge carriers. Rearranging Eq. (3) leads to

$$j_{\text{exc}} = qS_{\text{exc}} n_i \left[ \frac{n}{n_i} - \exp\left( \frac{qV_{\text{elec}}}{k_B T} \right) \right] = j_B \left[ \exp\left( \frac{qV_{\text{int}}}{k_B T} \right) - \exp\left( \frac{qV_{\text{elec}}}{k_B T} \right) \right]. \tag{4}$$

Thus, equation (4) is equivalent to the alternative equivalent circuit model of Breitenstein [21], which is exact for single sided pn-junctions as shown in Ref. [23]. The prefactor $j_B$ is a reference current density that is related to the carrier exchange velocity $S_{\text{exc}}$ controlling the efficacy of carrier exchange between the absorber and external terminal of the device.

The model of Refs. [21], [23] account for a single-sided pn-junction, where the total external voltage drops over this pn-junction and the internal voltage is given by the chemical potential of the minority carriers. In a more general case, we account for the fact that the electrical voltage drops over both contacts such that $V_{\text{elec}} = V_{\text{elec,n}} + V_{\text{elec,p}}$, likewise $V_{\text{int}} = V_{\text{int,n}} + V_{\text{int,p}}$.[35] Considering for a simple model of a PSC, a symmetric constellation (depicted in figure 2(a)), we arrive at

$$j_{\text{exc}} = j_B \left[ \exp\left( \frac{qV_{\text{int}}}{2k_B T} \right) - \exp\left( \frac{qV_{\text{elec}}}{2k_B T} \right) \right]. \tag{5}$$

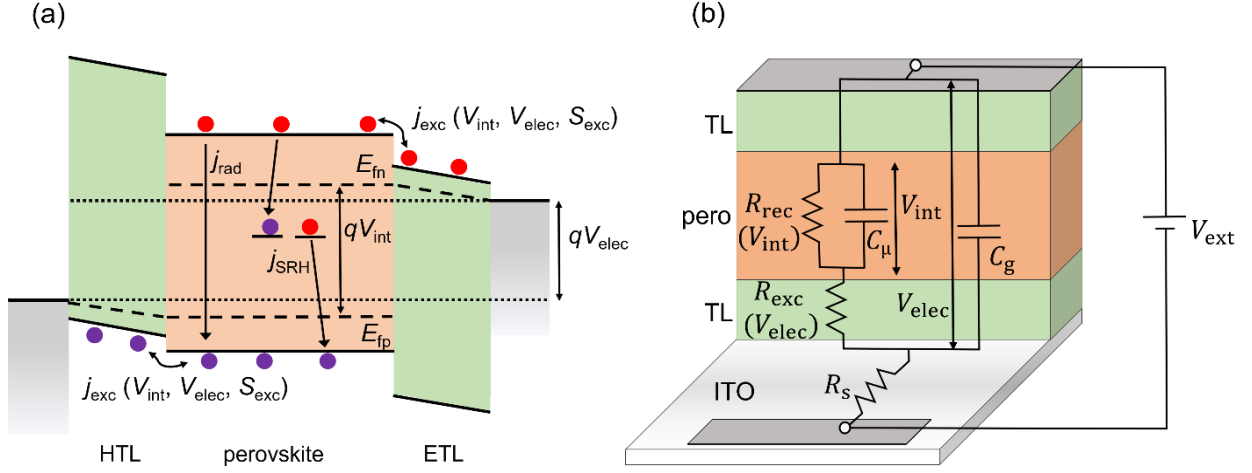

**Figure 2** (a) Schematic of physical mechanisms (represented as current densities $j$) that are considered in the matrix model. We consider a symmetric device with equal exchange velocity $S_{exc}$ for both transport layers. $j_{rad}$ is the radiative recombination current density and $j_{SRH}$ is the Shockley-Read-Hall non-radiative recombination current density. Electrons and holes are represented as red and purple spheres respectively. $V_{int}$ is the quasi-Fermi level splitting inside the perovskite layer and $V_{elec}$ is the quasi-Fermi level of the electrons at the cathode minus the quasi-Fermi level of the holes at the anode. (b) Equivalent circuit derived from the model (see section A4 in the SI for derivation). The exchange resistance $R_{exc}$ models the potential drop across both the transport layers (TL) that translates the quasi-Fermi level splitting in the perovskite $V_{int}$ to $V_{elec}$. $R_{exc}$ itself depends on $V_{elec}$ and the exchange velocity $S_{exc}$.

Note that with the assumption of symmetry between the behaviour of electrons and holes, the exchange current $j_{exc}$ is the same at the electron contact and at the hole contact (figure 2(a)). While the prefactor $j_B$ is voltage independent in pn-junctions,[23] it becomes voltage dependent in the case of undoped electron and hole transport layers over which a significant electric field drops. In case the electric field is approximately constant over the transport layers, an analytical solution for $j_B$ exists, given by

$$j_B = \frac{q\mu_{TL}F_{TL}n_i \exp\left(\frac{q\Phi_{b,pero}}{k_B T}\right)}{1-\exp\left(-\frac{qF_{TL}d_{TL}}{k_B T}\right)} = qn_{int}S_{exc}(V_{elec}), \tag{6}$$

which is consistent with the derivations found in ref.[36]. In equation 6, $\mu_{TL}$ is the mobility of electrons in the ETL and $n_{int} = n_i \exp(q\Phi_{b,pero}/k_B T)$ is the electron concentration at the interface between the perovskite and the ETL, with $n_i$ being the intrinsic concentration in the perovskite absorber and $q\Phi_{b,pero}$ the band offset for electrons at the perovskite/ETL interface. $S_{exc}$ is an exchange velocity, which - within the validity range of equation (6) - is given by[36]

$$S_{exc}(V_{elec}) = \frac{\mu_{TL}F_{TL}}{1-\exp\left(-\frac{qF_{TL}d_{TL}}{k_B T}\right)} . \tag{7}$$

$S_{exc}$ is a function of the mobility $\mu_{TL}$ and the electric field $F_{TL}$ in the transport layer, determining the magnitude of the gradient in electrochemical potential required to sustain the current flow through the device. The electric field $F_{TL}$ is given by

$$F_{TL} = \frac{(V_{bi,TL} - \frac{V_{elec}}{k})}{d_{TL}}, \tag{8}$$

where $V_{bi,TL}$ is the built-in electrostatic potential in the transport layer and $k$ is a factor that controls the amount of the electrode voltage that goes to the transport layer, i.e., $k = 2$ for the symmetric case (cf. figure 2(a)). We note that equations 5-6 are an alternate representation (see section A2 in the SI for derivation) of the formula derived in the SI of ref.[18].

*B. Charge Carrier Exchange between absorber and contact*

Now we turn to the overall balance of the charge carriers in the perovskite absorber. Upon assuming that the mobility of the electrons and holes is high enough, we may describe the total amount of electrons in the absorber by the concentration $n$ multiplied by the absorber thickness $d$. Thus, the balance of electrons in the absorber is given by photogeneration and recombination and by the extraction/injection of electrons to/from the contact according to

$$\frac{dn}{dt} = -\frac{j_{\text{exc}}}{qd} + \frac{j_\Phi}{qd} - \frac{n}{\tau_{\text{SRH}}} - B_{\text{rad}} n^2 , \tag{9}$$

where $j_\Phi$ is the input photon flux that corresponds to generation rate, $\tau_{\text{SRH}}$ is the lifetime corresponding to bulk Shockley-Read-Hall (SRH) recombination and $B_{\text{rad}}$ is the radiative recombination coefficient (cm$^3$/s). We have assumed equal electron and hole densities and negligible de-trapping rates to obtain a first-order SRH recombination rate with a lifetime $\tau_{\text{SRH}} = \tau_{\text{n}} + \tau_{\text{p}}$, where $\tau_{\text{n}}$ and $\tau_{\text{p}}$ are the SRH lifetimes of electrons and holes. A schematic of the different processes considered in the model is shown in figure 2(a). Note that, for simplicity, we have omitted here recombination at the interfaces between absorber and the transport layers. If required, interface recombination velocities $S_{\text{intrec}}$ could be easily introduced by defining an additional related lifetime $\tau_{\text{intrec}} = d_{\text{pero}}/S_{\text{intrec}}$.

Now we turn to the linearized small-signal analysis of Eq. (9), using small-perturbation quantities as symbols with a tilde. Applying a small perturbation to equations 5 and 9 and removing steady-state terms, we obtain

$$\frac{d\tilde{n}}{dt} = -\left[\frac{1}{\tau_{\text{SRH}}} + \frac{1}{\tau_{\text{rad}}}\right]\tilde{n} - \frac{1}{qd}\left[\frac{dj_{\text{exc}}}{dV_{\text{elec}}}\tilde{V}_{\text{elec}} + \frac{dj_{\text{exc}}}{d\tilde{n}}\tilde{n}\right] + \frac{\tilde{J}_\Phi}{qd}$$
$$= -\frac{\tilde{n}}{\tau_{\text{eff}}} - \frac{1}{qd}\left[\frac{\tilde{V}_{\text{elec}}}{R_{\text{exc}}} + S_{\text{exc}}\tilde{n}\right] + \frac{\tilde{J}_\Phi}{qd}, \tag{10}$$

where $\tau_{\text{eff}}$ is the effective recombination lifetime given by the parallel combination of the SRH lifetime and radiative lifetime $\tau_{\text{rad}} = 1/(2B_{\text{rad}}n)$. In order to rewrite Eq.(10) for the time derivative of the modulated internal voltage $\tilde{V}_{\text{int}}$, we use Eq. (1) to obtain

$$\tilde{n} = \left(\frac{dn}{dV_{\text{int}}}\right)\tilde{V}_{\text{int}} = \frac{qn_{\text{i}}}{2k_{\text{B}}T}\exp\left(\frac{qV_{\text{int}}}{2k_{\text{B}}T}\right)\tilde{V}_{\text{int}} = \frac{C_\mu}{qd}\tilde{V}_{\text{int}}, \tag{11}$$

with the definition of the chemical capacitance $C_\mu$ given by [37]

$$C_\mu = qd\left(\frac{dn}{dV_{\text{int}}}\right) = \frac{q^2 dn_{\text{i}}}{2k_{\text{B}}T}\exp\left(\frac{qV_{\text{int}}}{2k_{\text{B}}T}\right). \tag{12}$$

With this, we obtain

$$C_\mu \frac{d\tilde{V}_{\text{int}}}{dt} = -\frac{\tilde{V}_{\text{int}}}{R_{\text{rec}}} - \left[\frac{dj_{\text{exc}}}{dV_{\text{elec}}}\tilde{V}_{\text{elec}} + \frac{dj_{\text{exc}}}{dV_{\text{int}}}\tilde{V}_{\text{int}}\right] + \tilde{J}_\Phi. \tag{13}$$

Under open circuit bias condition, we have $V_{\text{elec}} = V_{\text{int}}$ and therefore

$$-\frac{dj_{\text{exc}}}{dV_{\text{elec}}} = \frac{dj_{\text{exc}}}{dV_{\text{int}}} = \frac{1}{R_{\text{exc}}}, \tag{14}$$

which implies for the exchange resistance

$$R_{\text{exc}} = \frac{2k_{\text{B}}T}{qj_{\text{B}}(V_{\text{elec}})}\exp\left(-\frac{qV_{\text{elec}}}{2k_{\text{B}}T}\right) = \frac{2k_{\text{B}}T}{qj_{\text{B}}(V_{\text{int}})}\exp\left(-\frac{qV_{\text{int}}}{2k_{\text{B}}T}\right). \tag{15}$$

Thus, Eq. (10) transforms into

$$C_\mu \frac{d\tilde{V}_{\text{int}}}{dt} = -\frac{\tilde{V}_{\text{int}}}{R_{\text{rec}}} + \frac{\tilde{V}_{\text{elec}}}{R_{\text{exc}}} - \frac{\tilde{V}_{\text{int}}}{R_{\text{exc}}} + \tilde{J}_\Phi \tag{16}$$

with the recombination resistance $R_{\text{rec}}$ given by

$$R_{\text{rec}} = \frac{C_\mu}{\left(\frac{1}{\tau_{\text{SRH}}}+\frac{1}{\tau_{\text{rad}}}\right)^{-1}} = \left[\frac{1}{\frac{2k_BT\tau_b}{q^2dn_i}\exp\left(-\frac{qV_{\text{int}}}{2k_BT}\right)} + \frac{1}{\frac{k_BT}{q^2dn_i^2B_{\text{rad}}}\exp\left(-\frac{qV_{\text{int}}}{k_BT}\right)}\right]^{-1}. \quad (17)$$

Finally, we consider the time development of the small signal external voltage $\tilde{V}_{\text{elec}}$, taking into account that this voltage builds up over a geometric capacitance $C_g$. The charge needed to build up this voltage stems from the current leaving the absorber and the electrical current $\tilde{j}_{\text{elec}}$ from the outer circuit leading to

$$C_g \frac{d\tilde{V}_{\text{elec}}}{dt} = -\frac{\tilde{V}_{\text{elec}}}{R_{\text{exc}}} + \frac{\tilde{V}_{\text{int}}}{R_{\text{exc}}} - \tilde{j}_{\text{elec}}. \quad (18)$$

*C. Matrix equations*

At this point it is convenient to put the time development of both voltages $\tilde{V}_{\text{int}}$ and $\tilde{V}_{\text{elec}}$ into a matrix equation

$$\begin{pmatrix} C_\mu & 0 \\ 0 & C_g \end{pmatrix} \begin{pmatrix} \frac{d}{dt}\tilde{V}_{\text{int}} \\ \frac{d}{dt}\tilde{V}_{\text{elec}} \end{pmatrix} = \begin{pmatrix} -\frac{1}{R_{\text{exc}}} - \frac{1}{R_{\text{rec}}} & \frac{1}{R_{\text{exc}}} \\ \frac{1}{R_{\text{exc}}} & -\frac{1}{R_{\text{exc}}} \end{pmatrix} \begin{pmatrix} \tilde{V}_{\text{int}} \\ \tilde{V}_{\text{elec}} \end{pmatrix} + \begin{pmatrix} \tilde{j}_\phi \\ \tilde{j}_{\text{elec}} \end{pmatrix}. \quad (19)$$

In Eq. (19), all small signal quantities depend on time. The excitation can be chosen to be $\tilde{j}_\phi = \delta(0)$ and $\tilde{j}_{\text{elec}} = 0$ in order to describe the transient of the photovoltage (IMVS) $\tilde{V}_{\text{elec}}(t)$ following a delta excitation by a light pulse. Similarly, we may calculate the time transient of the internal voltage $\tilde{V}_{\text{int}}$, i.e. of the chemical potential, in order to describe the transient photoluminescence (tr-PL), see also Ref.[35].

When turning to frequency domain measurements, we interpret Eq. (19) as an equation for complex-valued small signal quantities by replacing the time derivative with the angular frequency $\omega$ with the imaginary unit $i$, leading to

$$i\omega \begin{pmatrix} C_\mu & 0 \\ 0 & C_g \end{pmatrix} \begin{pmatrix} \tilde{V}_{\text{int}}(\omega) \\ \tilde{V}_{\text{elec}}(\omega) \end{pmatrix} = \begin{pmatrix} -\frac{1}{R_{\text{exc}}} - \frac{1}{R_{\text{rec}}} & \frac{1}{R_{\text{exc}}} \\ \frac{1}{R_{\text{exc}}} & -\frac{1}{R_{\text{exc}}} \end{pmatrix} \begin{pmatrix} \tilde{V}_{\text{int}}(\omega) \\ \tilde{V}_{\text{elec}}(\omega) \end{pmatrix} + \begin{pmatrix} \tilde{j}_\phi(\omega) \\ \tilde{j}_{\text{elec}}(\omega) \end{pmatrix}. \quad (20)$$

Analogously to the use of Eq. (19) in the time domain, Eq. (20) can be used to describe different experimental methods in the frequency domain. E.g. using a periodic excitation of the chemical potential by photogeneration $\tilde{j}_\phi = \tilde{j}_\phi(\omega)$ and keeping the sample at open circuit voltage ($\tilde{j}_{\text{elec}} = 0$) would yield the signal $\tilde{V}_{\text{elec}}(\omega)$ as a solution of Eq. (20), the intensity modulated photovoltage signal (IMVS), the Fourier transform of TPV. Likewise, keeping the electrical voltage at zero ($\tilde{V}_{\text{elec}} = 0$) and detecting instead $\tilde{j}_{\text{elec}} = \tilde{j}_{\text{elec}}(\omega)$ corresponds to the result of intensity modulated photocurrent spectroscopy (IMPS). We may also consider an excitation by electrical current $\tilde{j}_{\text{elec}} = \tilde{j}_{\text{elec}}(\omega)$ not making any periodic optical excitation ($\tilde{j}_\phi = 0$) and detect the response $\tilde{V}_{\text{elec}}(\omega)$ of the electrical voltage, in order to describe impedance spectroscopy (IS). However, for the two latter experimental methods we need to consider an additional series resistance $R_s$ in the complete equivalent circuit (shown in figure 2(b)), which slightly modifies the simple matrix equations (19) and (20).

The relation between the applied external voltage $V_{\text{ext}}$ and the voltage at the electrodes $V_{\text{elec}}$ is given using the external series resistance $R_s$ as

$$\tilde{V}_{\text{elec}} = \tilde{V}_{\text{ext}} + R_s \tilde{j}_{\text{elec}}. \quad (21)$$

Substituting equation 21 in equations 16 and 18, we obtain

$$\tilde{j}_\Phi + \frac{R_s}{R_{\text{exc}}}\tilde{j}_{\text{elec}} = \left[\frac{d}{dt} + \left(\frac{1}{R_{\text{exc}}} + \frac{1}{R_{\text{rec}}}\right)\right] C_\mu \tilde{V}_{\text{int}} - \frac{\tilde{V}_{\text{ext}}}{R_{\text{exc}}}, \quad (22)$$

$$\tilde{J}_{elec}\left(1 + \frac{R_s}{R_B} + C_g R_s \frac{d}{dt}\right) = \frac{\tilde{V}_{int}}{R_{exc}} - \left(\frac{1}{R_{exc}} + C_g \frac{d}{dt}\right)\tilde{V}_{ext}. \tag{23}$$

Combining the frequency domain versions ($d/dt \to i\omega$) of equations 22-23 and rearranging terms, we can obtain the matrix equations corresponding to different frequency domain techniques as

**IMVS**

$$i\omega \begin{pmatrix} C_\mu & 0 \\ 0 & C_g \end{pmatrix} \begin{pmatrix} \tilde{V}_{int}(\omega) \\ \tilde{V}_{ext}(\omega) \end{pmatrix} = \begin{pmatrix} -\frac{1}{R_{exc}} - \frac{1}{R_{rec}} & \frac{1}{R_{exc}} \\ \frac{1}{R_{exc}} & -\frac{1}{R_{exc}} \end{pmatrix} \begin{pmatrix} \tilde{V}_{int}(\omega) \\ \tilde{V}_{ext}(\omega) \end{pmatrix} + \begin{pmatrix} \tilde{J}_\Phi(\omega) \\ 0 \end{pmatrix}. \tag{24}$$

**IMPS**

$$i\omega \begin{pmatrix} C_\mu & 0 \\ 0 & C_g R_s \end{pmatrix} \begin{pmatrix} \tilde{V}_{int}(\omega) \\ \tilde{J}_{elec}(\omega) \end{pmatrix} = \begin{pmatrix} -\frac{1}{R_{exc}} - \frac{1}{R_{rec}} & \frac{R_s}{R_{exc}} \\ \frac{1}{R_{exc}} & -1 - \frac{R_s}{R_{exc}} \end{pmatrix} \begin{pmatrix} \tilde{V}_{int}(\omega) \\ \tilde{J}_{elec}(\omega) \end{pmatrix} + \begin{pmatrix} \tilde{J}_\Phi(\omega) \\ 0 \end{pmatrix}. \tag{25}$$

**IS**

In the case of IS, we require an additional arbitrary geometric capacitor $C_b$ that is placed before $R_s$, with a current $\tilde{J}_{term}$ flowing at the terminals. This yields a $3 \times 3$ matrix that is given by

$$i\omega \begin{pmatrix} C_\mu & 0 & 0 \\ 0 & C_g & C_g R_s \\ 0 & C_b & 0 \end{pmatrix} \begin{pmatrix} \tilde{V}_{int}(\omega) \\ \tilde{V}_{ext}(\omega) \\ \tilde{J}_{elec}(\omega) \end{pmatrix}$$
$$= \begin{pmatrix} -\frac{1}{R_{exc}} - \frac{1}{R_{rec}} & \frac{1}{R_{exc}} & \frac{R_s}{R_{exc}} \\ \frac{1}{R_{exc}} & -\frac{1}{R_{exc}} & -1 - \frac{R_s}{R_{exc}} \\ 0 & 0 & 1 \end{pmatrix} \begin{pmatrix} \tilde{V}_{int}(\omega) \\ \tilde{V}_{ext}(\omega) \\ \tilde{J}_{elec}(\omega) \end{pmatrix} + \begin{pmatrix} \tilde{J}_\Phi(\omega) \\ 0 \\ -\tilde{J}_{term}(\omega) \end{pmatrix}. \tag{26}$$

The solutions of matrices 24-26 and their corresponding solutions in the time domain are shown in table S2 in the SI. The n × n matrix yields eigenvalues $\lambda_1 \ldots \lambda_n$ that are connected to the time constants through $\lambda_n = 1/\tau_n$. While a simple analytic solution for the time constants of IS was not achievable, an analytic solution of the transfer function can be obtained using the relation $Z = W/Q$. In the case of frequency domain measurements, the expressed time constants are determined from the peak of the imaginary part of the solution of the transfer function (table S2 in the SI) versus frequency.

### D. Equivalent circuit

An alternate way of solving for the expressed time constants is to use equations 22 and 23 to calculate the IMVS, IMPS and IS transfer functions, followed by obtaining the expressed time constant through the peak of the imaginary part of the transfer function versus frequency. These transfer functions are given by (derivation in section A4 in the SI)

$$W = \frac{\tilde{V}_{ext}}{\tilde{J}_\Phi} = \left(\frac{1}{R_{rec}} + i\omega C_\mu + i\omega C_g \left(1 + \frac{R_{exc}}{R_{rec}} + i\omega R_{exc} C_\mu\right)\right)^{-1}, \tag{27}$$

$$Q = \frac{\tilde{J}_{\text{elec}}}{\tilde{J}_\Phi} = \left(1 + \frac{R_s + R_{\text{exc}}}{R_{\text{rec}}} + i\omega(R_s + R_{\text{exc}})C_\mu + i\omega R_s C_g\left(1 + \frac{R_{\text{exc}}}{R_{\text{rec}}} + i\omega R_{\text{exc}}C_\mu\right)\right)^{-1},$$
(28)

$$Z = \frac{\tilde{V}_{\text{ext}}}{\tilde{J}_{\text{elec}}} = R_s + \left(\frac{1}{R_{\text{exc}}+\left(\frac{1}{R_{\text{rec}}}+i\omega C_\mu\right)^{-1}} + i\omega C_g\right)^{-1} = \frac{W}{Q}.$$
(29)

Equations 27-29 yield the equivalent circuit of figure 2(b). This circuit contains the traditional elements of the solar cell model with the parallel combination of the recombination resistance and chemical capacitance.[38] The novel aspect of the model is the voltage-dependent exchange resistance $R_{\text{exc}}$ (that is in series with the $R_{\text{rec}}C_\mu$ line) which models the potential drop across the transport layers. Equations 6,7 and 15 show that $R_{\text{exc}}$ depends on the steady-state quasi-Fermi level splitting at the electrodes $V_{\text{elec}}$ and the mobility and electric field through the transport layer. Equations 27 and 28 identify a time constant called the exchange lifetime, given by

$$\tau_{\text{exc}} = R_{\text{exc}}C_\mu = \frac{d}{S_{\text{exc}}}.$$
(30)

The exchange lifetime determines how quickly charge carriers are extracted from (injected into) the volume of the absorber into (from) the contacts through the transport layers. We note that the resistance $R_{\text{exc}}$ is defined only under small perturbation conditions and is different to the steady-state transport layer resistance derived in our previous work.[18] The generation term is in the form of a current source $j_\Phi$ that is across the bulk $R_{\text{rec}}C_\mu$ line corresponding to the internal voltage $V_{\text{int}}$. The model is completed with the geometric capacitance $C_g$, which is in parallel to all the elements that are outside the external series resistance $R_s$.

### E. Model analysis

We now proceed to analyse the solution of the model. The parameters used to carry out the simulations are shown in table S3 in the SI. We consider two cases, one where the extraction lifetime is much faster than the effective recombination lifetime ($\tau_{\text{exc}} \ll \tau_{\text{eff}}$) and the other where the effective recombination lifetime is much faster than the extraction lifetime ($\tau_{\text{exc}} \gg \tau_{\text{eff}}$). A summary of these solutions are shown in figures 3(a,b) for the case $\tau_{\text{exc}} \ll \tau_b$, with the case $\tau_{\text{exc}} \gg \tau_b$ shown in the corresponding lower panels (figures 3(c,d)). We note that the time constants for IS and IMVS are identical, with the only difference being the voltage-independent third time constant $\tau = R_s C_b$ (not plotted in the graphs) for IS that attenuates the total current response. We therefore plot the IMVS time constants in the SI (figure S4 in the SI) and show only the time constants of IS and IMPS in our analysis.

In all cases, the model yields two time constants that correspond to each of the inverse eigenvalues, which are classified as a slow ($\tau_{\text{slow}}$) or a fast ($\tau_{\text{fast}}$) time constant with respect to each other. These eigenvalues are combined with their eigenvectors to yield the full solution of the transfer function (table S2 in the SI). The actual, expressed time constant is calculated from the frequency maxima of the imaginary part of the respective solution, simulated using the analytical solutions, termed 'solution' in the simulations in figure 3. An alternate way of calculating the expressed time constant or constants is to calculate the frequency maxima of the imaginary part of the transfer function via the equivalent circuit of figure 2(b), termed 'EC' in figure 3.

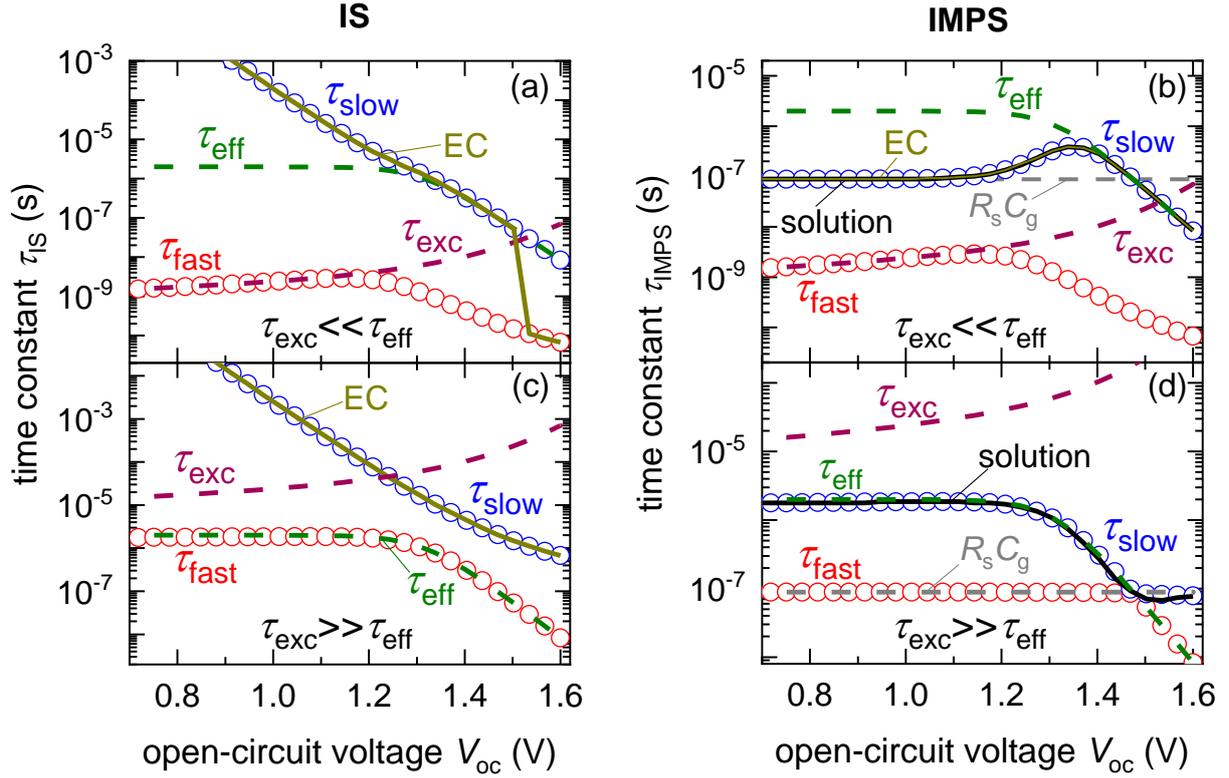

**Figure 3** Simulated analytical time constants $\tau_{\text{slow}}$ and $\tau_{\text{fast}}$ from the matrix model for $\tau_{\text{exc}} \ll \tau_{\text{eff}}$ (a, b) and $\tau_{\text{exc}} \gg \tau_{\text{eff}}$ (c, d), for IS (a, c) and IMPS (b, d). The third time constant of IS that attenuates the modulated current ($\tau = R_s C_b$) is not shown. The IMVS time constants are identical to that of IS, shown in figure S4 in the SI. 'Solution' and 'EC' correspond to the time constant calculated from the frequency maxima of the imaginary part of the analytical solution of the transfer function (see table S2 in the SI) and the equivalent circuit respectively. $\tau_{\text{eff}}$ is the effective recombination lifetime consisting of the parallel combination of SRH and radiative recombination, and $\tau_{\text{exc}}$ is the exchange lifetime between the perovskite absorber and the contacts (equation 30). The expressed time constant in measurements is always the slower time constant $\tau_{\text{slow}}$. Simulation parameters are shown in table S3 in the SI. The equality of the EC solution and the analytical solution of IS is shown in figure S5 in the SI.

For the case of fast charge carrier exchange ($\tau_{\text{exc}} \ll \tau_{\text{eff}}$, figure 3(a, b)), the IS and IMPS solutions yield a slow time constant that corresponds to the effective recombination lifetime $\tau_{\text{eff}}$ at large open-circuit voltages. $\tau_{\text{eff}}$ is voltage-dependent due to the dominance of radiative recombination over SRH recombination at high open-circuit voltages (equations 17). In the absence of radiative recombination, $\tau_{\text{eff}}$ will correspond to the constant SRH lifetime, which forms a plateau in the time constant. This region then transitions into an exponentially increasing time constant at lower open-circuit voltages. For IMPS, the slow time constant corresponds to the product of the series resistance $R_s$ and the geometric capacitance $C_g$ at lower open-circuit voltages (referred to as $RC$ attenuation in literature[39,40]) and transitions to $\tau_{\text{eff}}$ at higher open-circuit voltages. The fast time constant for all the transfer functions corresponds to the exchange lifetime $\tau_{\text{exc}}$ at lower open-circuit voltages before transitioning to an exponential reduction at higher open-circuit voltages. The open-circuit voltage dependence of $\tau_{\text{exc}}$ arises out of equations 7 and 8 with $V_{\text{elec}} = V_{\text{oc}}$, causing it to increase at higher open-circuit voltages.

For the case of slow charge carrier exchange ($\tau_{\text{exc}} \gg \tau_{\text{eff}}$, figure 3(c, d)), in the case of IS, the fast time constant corresponds to $\tau_{\text{eff}}$. The slow time constant does not follow $\tau_{\text{exc}}$ but shows a much larger magnitude at low open-circuit voltages, decreasing exponentially followed by a saturation at high open-circuit voltages. In the case of IMPS, the fast time constant corresponds to the $R_s C_g$ product, while the slow time constant corresponds to $\tau_{\text{eff}}$. At large open-circuit voltages, the IMPS fast and slow time constants are exchanged, which results in the time constant values not crossing each other. We highlight that for all the transfer functions in both cases, the expressed time constant from the solution of the model and the equivalent circuit corresponds to the slower time constant. Therefore, when $\tau_{\text{exc}} \ll \tau_{\text{eff}}$, which is the case expected for a high efficiency PSC with good charge extraction from the perovskite to the contacts through the transport layers, the faster time constant that contains information on the exchange properties is not directly accessible.

We now aim to identify the different unknown contributions to the slow and fast time constants, shown in figure 4. For the case of fast charge carrier exchange ($\tau_{\text{exc}} \ll \tau_{\text{eff}}$), the exponential drop in the fast time constant at high open-circuit voltages for all the transfer functions is determined by the product $R_{\text{exc}} C_g$. Furthermore, the exponential rise in the slow (expressed) time constant for IS (figure 4(a)) at low open-circuit voltages corresponds to the product of the recombination resistance $R_{\text{rec}}$ and the geometric capacitance $C_g$. This coupling can also be seen from the experimental resistances and capacitances (figures 1(b) and 1(c) respectively) at low open-circuit voltages, where the resistance increases exponentially while the capacitance has saturated to a constant value. This effect was labelled as the recombination of electrode charges by Kiermasch et al.[8], who identified this effect for different photovoltaic technologies when observing large lifetimes in the order of milliseconds to seconds from TPV and open-circuit voltage decay measurements. The same phenomenon was later reported on by Wang et al.[28] in the case of TPV measurements and IS simulations, who concluded that the lifetimes observed were not representative of recombination dynamics but dominated by the geometric capacitance of the device. Krückemeier et al.[7] also compared the reported lifetimes from TPV and tr-PL measurements, concluding that the much larger lifetimes reported from TPV measurements were due to measurement of this $R_{\text{rec}} C_g$ time constant. This occurs because TPV measurements require a full device with electrodes that introduces the effects of the geometric capacitance, whereas tr-PL is mostly measured on device stacks without electrodes.

For the case of slow charge carrier exchange ($\tau_{\text{exc}} \gg \tau_{\text{eff}}$), the slow time constant in the case of IS is determined by the product of $R_{\text{exc}}$ and $C_g$. The voltage dependence of the extraction velocity $S_{\text{exc}}$ (equation 7) creates a unique evolution of $R_{\text{exc}}$ (equation 15) and hence the $R_{\text{exc}} C_g$ time constant, which shows an exponential decrease followed by a saturation at high open-circuit voltages, which could be misinterpreted as an apparent SRH lifetime. Since the slow time constant is the one that is expressed during a measurement, information regarding the transport layer can be obtained through the determination of $R_{\text{exc}}$ as a function of the open-circuit voltage in this case. An intermediate case is shown in figure S6 in the SI for IS measurements, when $\tau_{\text{exc}} \cong \tau_{\text{eff}}$. In this case, the resistance at large open-circuit voltages is dominated by $R_{\text{exc}}$ while at lower open-circuit voltages, $R_{\text{rec}}$ dominates. This situation leads to the slow (expressed) time constant at large open-circuit voltages being dominated by the $R_{\text{exc}} C_g$ product that makes a plateau, while at lower open-circuit voltages, it is dominated by the product $R_{\text{rec}} C_g$.

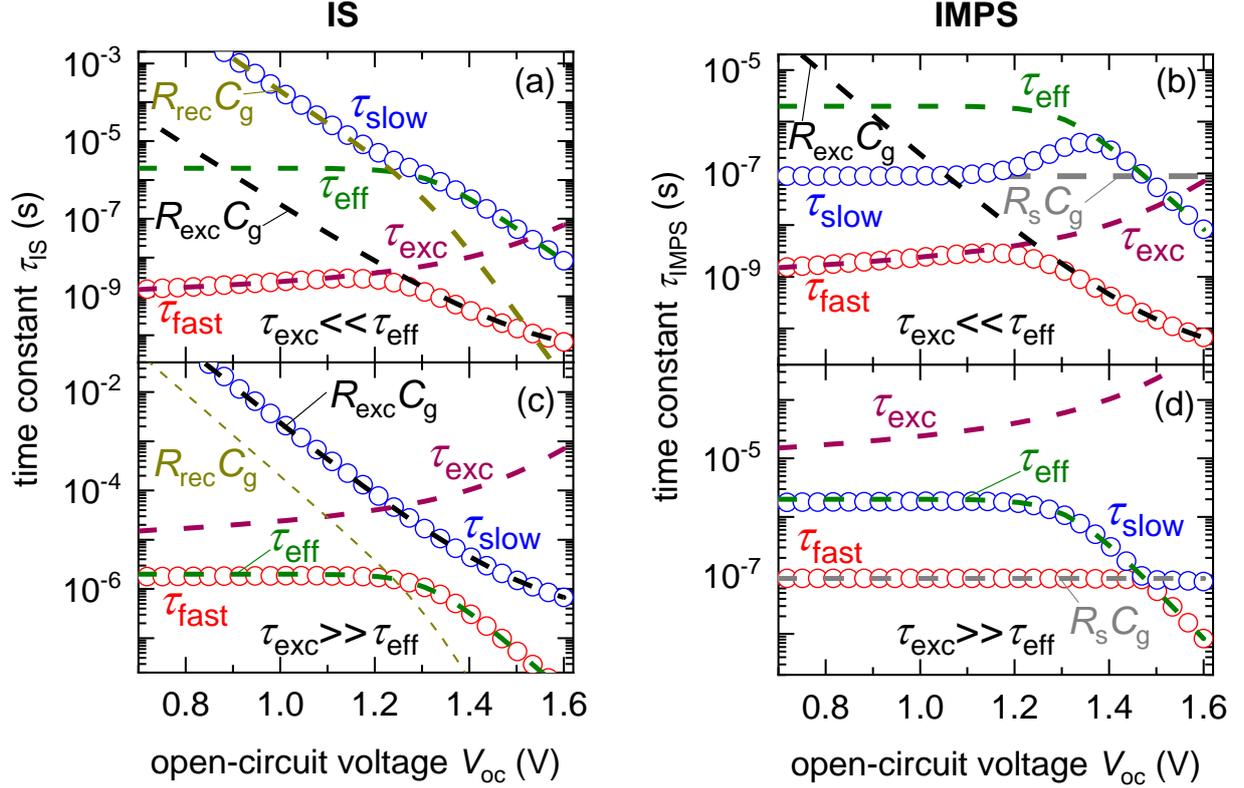

**Figure 4** Simulated analytical time constants $\tau_{\text{slow}}$ and $\tau_{\text{fast}}$ from the matrix model for $\tau_{\text{exc}} \ll \tau_{\text{eff}}$ (a,b) and $\tau_{\text{exc}} \gg \tau_{\text{eff}}$ (c,d), for IS (a,c) and IMPS (b,d). The IMVS time constants are identical to that of IS, shown in figure S4 in the SI. The different contributions to the time constants are shown with the corresponding labels. $R_{\text{rec}}$ is the recombination resistance (equation 17) while $R_{\text{exc}}$ is the exchange resistance (equation 15). For IS measurements, $R_{\text{rec}}C_g$ dominates the expressed time constant ($\tau_{\text{slow}}$) at lower open-circuit voltages when $\tau_{\text{exc}} \ll \tau_{\text{eff}}$, while $R_{\text{exc}}C_g$ dominates the expressed time constant when $\tau_{\text{exc}} \gg \tau_{\text{eff}}$. $\tau_{\text{fast}}$ for IS is given by $\tau_{\text{exc}}$ at low open-circuit voltages followed by $R_{\text{exc}}C_g$ at high open-circuit voltages when $\tau_{\text{exc}} \ll \tau_{\text{eff}}$, while $\tau_{\text{fast}} = \tau_{\text{eff}}$ for the case $\tau_{\text{exc}} \ll \tau_{\text{eff}}$.

To provide an intuitive connection between the analysis of the time constants in figures 3,4 and the equivalent circuit in figure 2(b), we calculate the effective resistance $R_{\text{eff}}$ and capacitance $C_{\text{eff}}$ for the two limiting cases. For the case $\tau_{\text{exc}} \ll \tau_{\text{eff}}$, when $\omega^2 \tau_{\text{eff}}^2 \ll 1$, for $\tau_{\text{exc}} \ll \tau_{\text{eff}}$, we have (from equation 29)

$$Z(\tau_{\text{exc}} \ll \tau_{\text{eff}}, C_\mu \gg C_g) = R_s + R_{\text{exc}} + \left(\frac{1}{R_{\text{rec}}} + i\omega C_\mu\right)^{-1}. \tag{31}$$

The effective low-frequency resistance is

$$R_{\text{eff}}(\tau_{\text{exc}} \ll \tau_{\text{eff}}, C_\mu \gg C_g) = R_s + R_{\text{exc}} + R_{\text{rec}}. \tag{32}$$

For the case $C_\mu \ll C_g$, we have

$$Z(\tau_{\text{exc}} \ll \tau_{\text{eff}}, C_\mu \ll C_g) = R_s + \left(\frac{1}{R_{\text{rec}}} + i\omega C_g\right)^{-1}, \tag{33}$$

with the effective low-frequency resistance being

$$R_{\text{eff}}(\tau_{\text{exc}} \ll \tau_{\text{eff}}, C_\mu \ll C_g) = R_s + R_{\text{rec}}. \tag{34}$$

The effective low-frequency capacitance for both cases is

$$C_{\text{eff}}(\tau_{\text{exc}} \ll \tau_{\text{eff}}) = \left(\frac{R_{\text{rec}}}{R_{\text{eff}}}\right)^2 C_{\text{pero,eff}}, \tag{35}$$

where $C_{\text{pero,eff}} = C_\mu$ and $R_{\text{eff}} = R_s + R_{\text{exc}} + R_{\text{rec}}$ when $C_\mu \gg C_g$, $C_{\text{pero,eff}} = C_g$ and $R_{\text{eff}} = R_s + R_{\text{rec}}$ when $C_\mu \ll C_g$. For the case $\tau_{\text{exc}} \gg \tau_{\text{eff}}$, when $\omega^2 \tau_{\text{exc}}^2 \ll 1$, we have

$$Z(\tau_{\text{exc}} \gg \tau_{\text{eff}}) = R_s + (\frac{1}{R_{\text{exc}}} + i\omega C_g)^{-1}. \tag{36}$$

The effective low-frequency resistance is

$$R_{\text{eff}}(\tau_{\text{exc}} \gg \tau_{\text{eff}}) = R_s + R_{\text{exc}} \tag{37}$$

and the effective low-frequency capacitance is

$$C_{\text{eff}}(\tau_{\text{exc}} \gg \tau_{\text{eff}}) = (\frac{R_{\text{exc}}}{R_s + R_{\text{exc}}})^2 C_g. \tag{38}$$

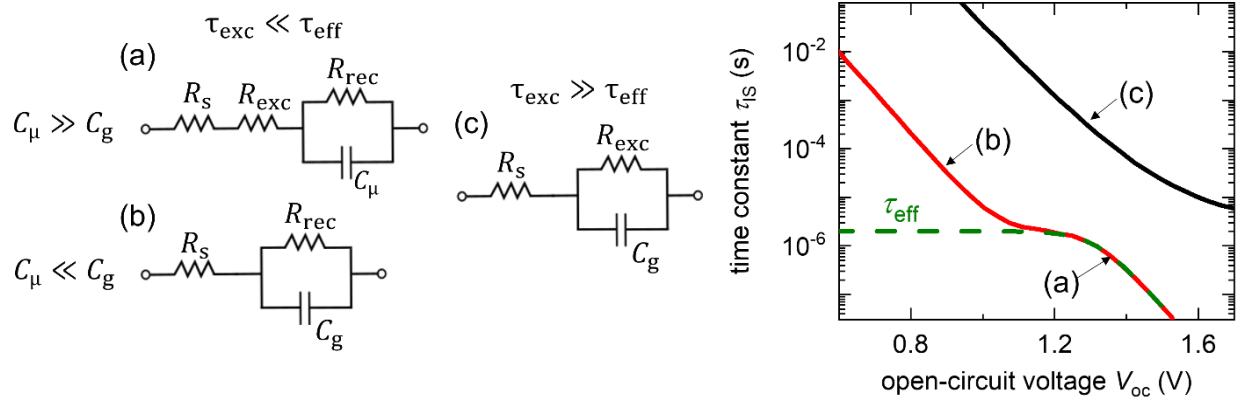

**Figure 5** Reduced equivalent circuit derived from the matrix model (see figure 2(b)) for the special cases (a,b) $\tau_{\text{exc}} \ll \tau_{\text{eff}}$ and (c) $\tau_{\text{exc}} \gg \tau_{\text{eff}}$. (a) corresponds to the case when the chemical capacitance $C_\mu$ is much larger than the geometric capacitance $C_g$ (large forward bias), while (b) corresponds to $C_\mu \ll C_g$ (reverse bias and low forward bias). The $R\|C$ products that form the time constant in each case ($\tau = \tau_{\text{eff}} = R_{\text{rec}}C_\mu$ in (a), $\tau = R_{\text{rec}}C_g$ in (b) and $\tau = R_{\text{exc}}C_g$ in (c)) are the expressed time constants in an IS measurement as shown in the panel on the right. The total resistance of the PSC is determined by the sum of the series resistance $R_s$, exchange resistance $R_{\text{exc}}$ and recombination resistance $R_{\text{rec}}$ (i.e. they are in series).

Figure 5 shows a summary of the reduced equivalent circuits for each case. When the exchange lifetime is much faster than the effective recombination lifetime ($\tau_{\text{exc}} \ll \tau_{\text{eff}}$), there are two possible scenarios. At large applied voltages, when the chemical capacitance is much larger than the geometric capacitance, the effective resistance is simply a series connection of the series resistance $R_s$, exchange resistance $R_{\text{exc}}$ and recombination resistance $R_{\text{rec}}$, while the time constant is given by the bulk recombination lifetime $\tau_{\text{eff}} = R_{\text{rec}}C_\mu$. At low applied voltages, when the chemical capacitance is smaller than the geometric capacitance, the effective resistance is dominated by the series resistance and the recombination resistance, while the time constant is given by $\tau = R_{\text{rec}}C_g$. These voltage-dependent time constants are also seen in figure 4(a). When the exchange lifetime is much slower than the effective recombination lifetime ($\tau_{\text{exc}} \gg \tau_{\text{eff}}$), the total resistance is a series connection of $R_s$ and $R_{\text{exc}}$, with the time constant given by $\tau_b = R_{\text{exc}}C_g$, also predicted in figure 4(c). The effective capacitance in all cases depends not only on the individual capacitances but also on the resistances (equations 35 and 37).[18] This effect is seen in figure S7 in the SI, where both the series resistance and exchange resistance create a saturation of the effective capacitance calculated from the matrix model (also discussed in ref.[41]).

## 4. Simulations

To confirm the validity of our model, we carry out IS simulations on a *p-i-n* PSC with the structure *ITO/PTAA/perovskite/PCBM/Ag*, whose band diagram is shown in figure 6(a). For simplicity, we consider no band offsets between the perovskite and transport layers, mobilities of electrons and holes to be equal in the respective layers and no external series resistance ($R_s = 0$). We consider a uniform volume density of mid-gap electron (acceptor-like) traps $N_t$ through the perovskite layer. Further details of simulation parameters are shown in table S4 in the SI. We also neglect ionic densities within the perovskite layer since we are interested only in electronic phenomena and also because the ionic contributions to the PSC spectra are known to occur below ~$10^3$ Hz in IS measurements,[24] indicating that the response at frequencies above this value are purely electronic in nature and hence can be represented by a purely electronic model. The strategy of this section is to simulate situations of increasing complexity, starting from equal mobilities of the perovskite and transport layers up to the situation of a full device with varying mobilities of the transport layers, in order to get a clear picture of the physical mechanisms that contribute to the response.

*Perfectly conducting transport layers, no bulk electric field*

We begin by considering equal mobilities of the perovskite layer ($\mu_{\text{pero}} = 20$ cm$^2$/Vs) and the transport layers. We further make the relative permittivity of the perovskite very high ($\varepsilon_{\text{r,pero}} = 3 \times 10^5$) to ensure that there is no electric field in the perovskite layer. Figure 6(b) shows the time constants calculated from the maxima of the imaginary part of the simulated IS transfer function for different SRH lifetimes. These time constants decrease exponentially with increasing open-circuit voltage with a subsequent change in slope that leads to approaching the same value (~$10^{-7}$ s) at high open-circuit voltages. For the same open-circuit voltage, a larger SRH lifetime corresponds to a larger magnitude of the time constant. Figures 6(c) and 6(d) show the resistance and capacitance respectively, whose product yields the measured time constants. We define a factor $m$ that controls the slope of the exponential evolution of the resistances $R$ and capacitances $C$ with respect to the open-circuit voltage, given by

$$R \propto \exp\left(-\frac{qV_{\text{oc}}}{m_R k_B T}\right), \tag{39}$$

$$C \propto \exp\left(\frac{qV_{\text{oc}}}{m_C k_B T}\right). \tag{40}$$

The slope factors of the resistances and capacitances in figure 6 are shown in figure S8 in the SI. The capacitance evolves identically for all cases, transitioning from the geometric capacitance at low open-circuit voltages to an exponential increase with a slope factor $m_C$ that approaches a value of 2 at high open-circuit voltages, corresponding to the chemical capacitance of an intrinsic layer (equation 12). The resistances show a slope factor $m_R$ equal to 2 at low open-circuit voltages, corresponding to the dominance of first-order SRH recombination in the bulk (equation 17). For higher open-circuit voltages, $m_R$ transitions to a value of 1 corresponding to the dominance of second-order radiative recombination. Therefore, the product of the resistances and capacitances in the low open-circuit voltage regime shows an exponential evolution ($\tau = R_{\text{rec,SRH}} C_g \propto \exp(-qV_{\text{oc}}/2k_B T)$) while in the high open-circuit voltage regime, the exponential trend is maintained due to their differing slopes ($\tau = \tau_{\text{rad}} \propto \exp(-qV_{\text{oc}}/2k_B T)$), leading to the time constants not saturating anywhere near their respective SRH lifetimes. An exception is the case of the smallest SRH lifetime (red stars in figure 6(b)) due to its slower transition of $m_R$ from a value of 2 compared to the other two cases. We thus speculate that the onset of radiative recombination prevents the observation of a plateau that corresponds to the SRH lifetime, which is also observed in the simulations from the matrix model shown in figure 3(a).

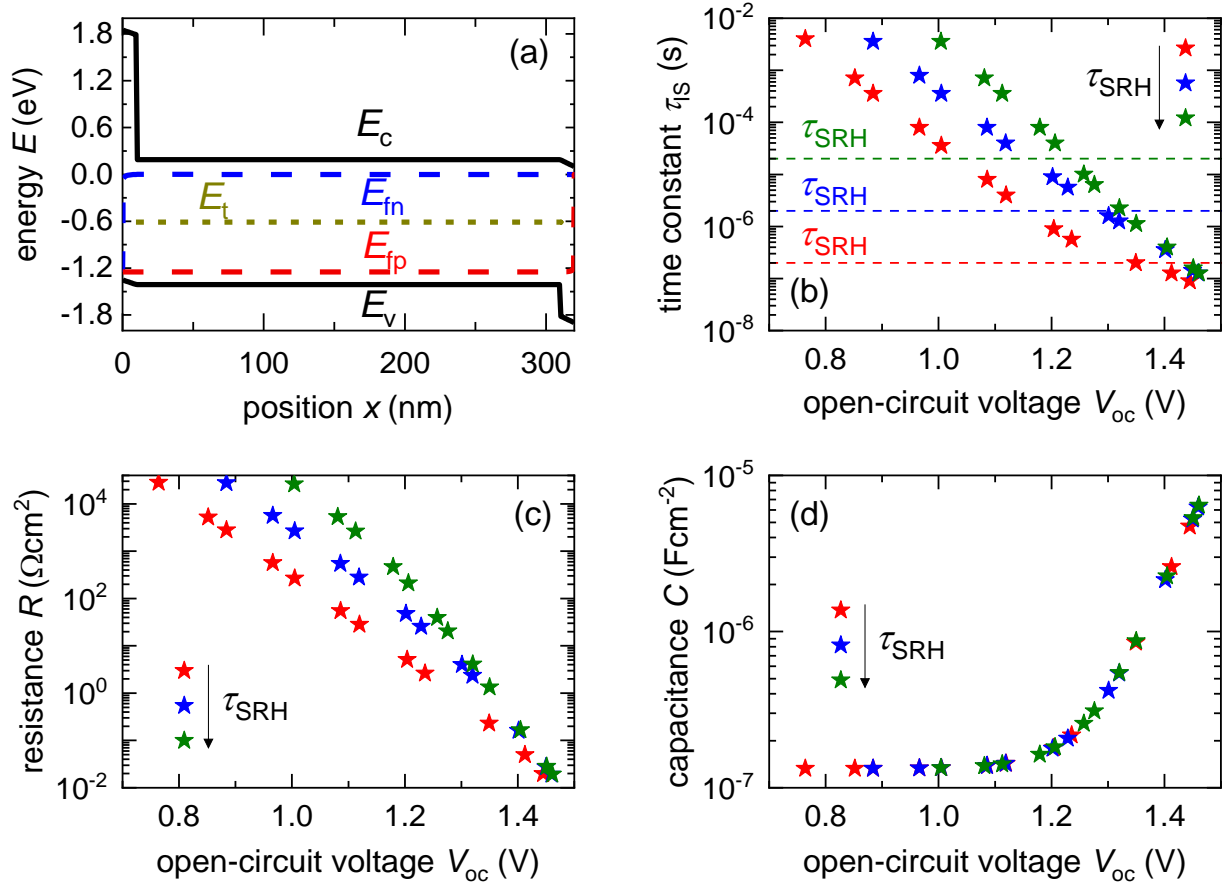

**Figure 6** (a) Band diagram of the *ITO/PTAA*/perovskite/*PCBM/Ag* PSC upon which the drift-diffusion simulations are based on. The relative permittivity of the perovskite was increased arbitrarily high to ensure no electric field within the perovskite layer. The mobilities of the transport layers were made equal to that of the perovskite layer. Calculated (a) high frequency time constants and corresponding (b) resistances and (c) capacitances from IS simulations at open-circuit for different SRH lifetimes (dotted lines in (b)). Simulation parameters are shown in table S4 in the SI. Radiative recombination prevents the observation of the plateau in the time constant corresponding to SRH recombination.

*Perfectly conducting transport layers, no bulk electric field, no radiative recombination*

To confirm our hypothesis, we switch off radiative recombination and redo the simulations of the previous section, shown in figure 7. We find that the time constants (symbols) now saturate nicely at the respective SRH lifetime at high open-circuit voltages. The slope factors of the resistances and capacitances are shown in figure S9 in the SI. The capacitances (symbols in figure 7(c)) again evolve identically in all cases, with slope factor ~2 at high open-circuit voltages corresponding to the chemical capacitance $C_\mu$, and the geometric capacitance $C_g$ expressed at low open-circuit voltages. The resistances (symbols in figure 7(b)) also show a slope factor close to 2, indicating a first-order SRH recombination resistance through the entire voltage range. Therefore, at low open-circuit voltages, we again observe the exponential $C_g$-dominated lifetime. At high open-circuit voltages, when the capacitance transitions from the $C_g$ to $C_\mu$, the product of the SRH recombination resistance and $C_\mu$ yields the constant SRH lifetime. Furthermore, the predicted time constants, resistances and capacitances from the model (lines in figures 7(a)-(c)) match nicely with the simulated data. The predicted saturation of the capacitance at low voltages $C_{sat}$ from the model is different from that of the simulations because the $C_g$ of the transport layers is not included in the model ($C_{sat} =$

$[1/C_{g,\text{ETL}} + 1/C_{g,\text{pero}} + 1/C_{g,\text{HTL}}]^{-1}$)[18].

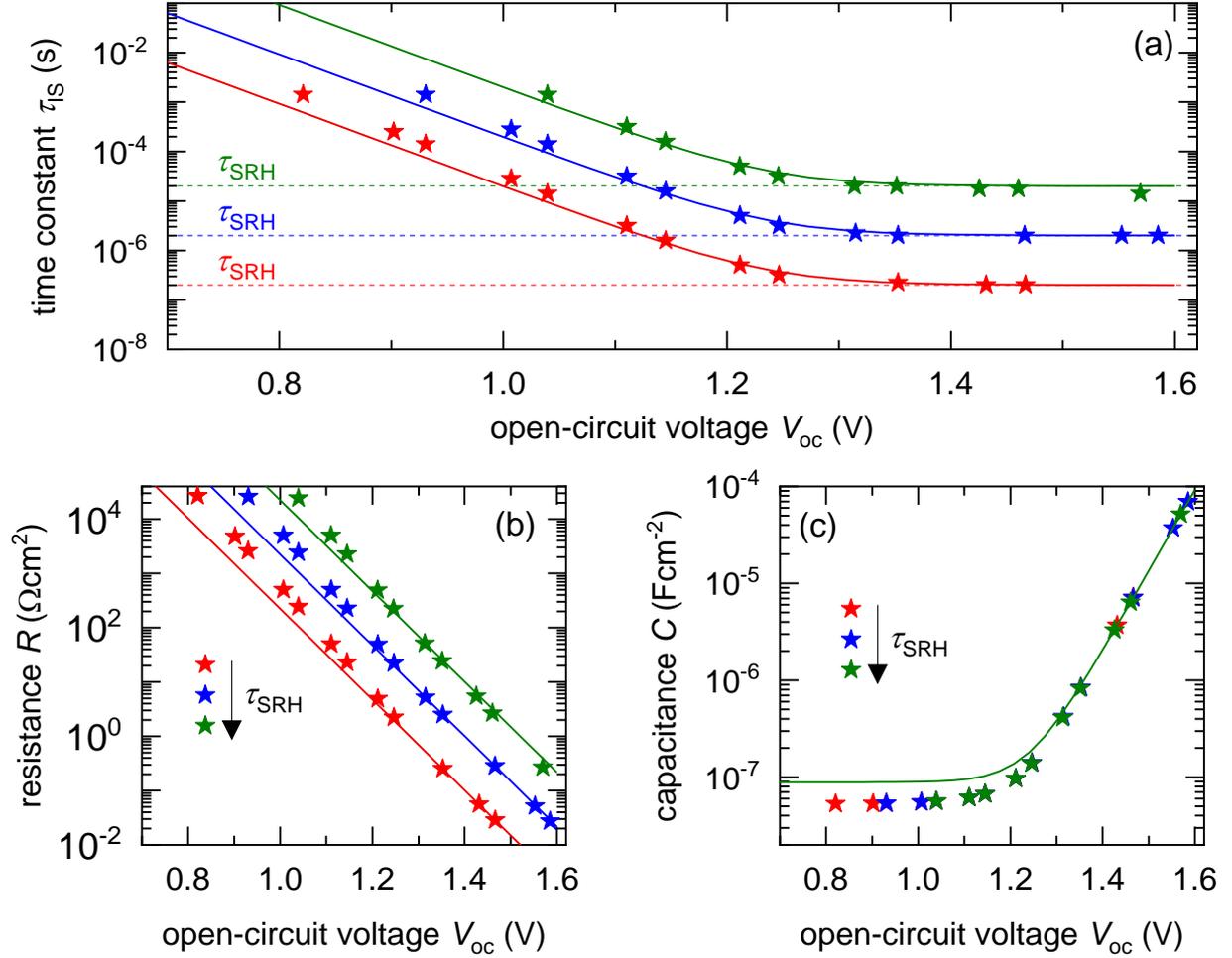

**Figure 7** Calculated (symbols) (a) high frequency time constants and corresponding (b) resistances and (c) capacitances from IS simulations at open-circuit for different SRH lifetimes (dotted lines in (a)) from drift-diffusion simulations of the PSC. In addition to ensuring no electric field in the perovskite layer with an arbitrarily high relative permittivity, radiative recombination was switched off for the perovskite layer. The mobilities of the transport layers were set equal to that of the perovskite layer. Simulation parameters are shown in table S4 in the SI. The lines in (a)-(c) are the corresponding predictions from the model. The absence of radiative recombination allows observation of a plateau in the time constant which corresponds to the SRH lifetime. The saturation of the capacitance at low open-circuit voltages predicted by the matrix model is different from the simulations because the model does not consider the individual geometric capacitances of the transport layers.

*Perfectly conducting HTL, variable ETL mobility*

We now increase the complexity of the simulations further, by varying the mobility of the ETL while keeping the HTL mobility equal to that of the perovskite layer. The simulated time constants are shown in figure 8(a), with the corresponding resistances and capacitances shown in figures 8(b) and 8(c) respectively. Quantities corresponding to different mobilities of the ETL are represented using both the symbols as shown in the legend and varying line darkness, while different SRH lifetimes are represented using different colours. The time constants show the same exponential evolution at low open-circuit voltages due to the dominance of $C_g$, as discussed in previous cases (figures 6 and 7), irrespective of the ETL mobility. At higher open-circuit voltages, this transitions into a plateau that shows a weak evolution versus open-circuit voltage and is strongly dependent on the ETL mobility but independent of the SRH lifetime. For lower mobilities, the time constant saturates at a corresponding larger value. Therefore, we conclude that the plateau region is not a real SRH lifetime in this case.

The corresponding resistances in figure 8(b) show two regimes, the first being an exponential evolution at low open-circuit voltages followed by a transition to a plateau at high open-circuit voltages, similar to the time constants. For higher mobilities of the ETL, we expect that, based on the findings of the previous sections, the exponential evolution at low open-circuit voltages is dominated by SRH recombination ($m_R = 2$). At higher open-circuit voltages, this region should transition into a steeper exponential drop due to the dominance of radiative recombination ($m_R = 1$). However, for lower ETL mobilities, the resistance makes a plateau at high open-circuit voltages. Simulations of the evolution of the total resistance versus transport layer mobility and open-circuit voltage from the model (figure S10 in the SI) show the same trends as observed in figure 8(b), where the larger resistance between $R_{\text{exc}}$ and $R_{\text{rec}}$ dominates the measured resistance (i.e. $R_{\text{exc}}$ and $R_{\text{rec}}$ are in series, as shown in figure 5). Therefore, for lower mobilities of the ETL, the expressed resistance is indeed $R_{\text{exc}}$, which transitions into the $R_{\text{rec}}$-dominated resistance for high ETL mobilities. The transition from $R_{\text{rec}}$ to $R_{\text{exc}}$ for reducing mobilities of the ETL is difficult to identify since both the resistances possess the same slope factor $m_R = 2$ (see equations 15 and 17) at low open-circuit voltages. The capacitance in figure 8(c) shows the typical geometric capacitance to chemical capacitance transition for high mobilities of the ETL. This transition is strongly dampened for lower ETL mobilities (larger $R_{\text{exc}}$), causing the capacitance to saturate at higher open-circuit voltages. This effect of $R_{\text{exc}}$ on the effective capacitance is also predicted by the model, shown in figure S7 in the SI.

To confirm the validity of our interpretation, we switch off radiative recombination and repeat the simulations, whose results are shown in figure S11 in the SI. For the case of highest mobility of the ETL ($\mu_{\text{ETL}} = \mu_{\text{pero}}$, represented using stars), the time constants saturate at their respective SRH lifetimes, while also observing the exponential $R_{\text{rec}}$ through the entire range of open-circuit voltage and the chemical capacitance at large open-circuit voltages. For lower mobilities of the ETL, the SRH lifetimes are not expressed anymore, instead replaced by the plateaus that are a function of the ETL mobility, with the corresponding dominance of $R_{\text{exc}}$ observed in the resistances. In summary, we conclude that for low mobilities of the transport layer, the plateaus in the time constants are not representative of the SRH lifetimes but are a consequence of the exchange resistance $R_{\text{exc}}$.

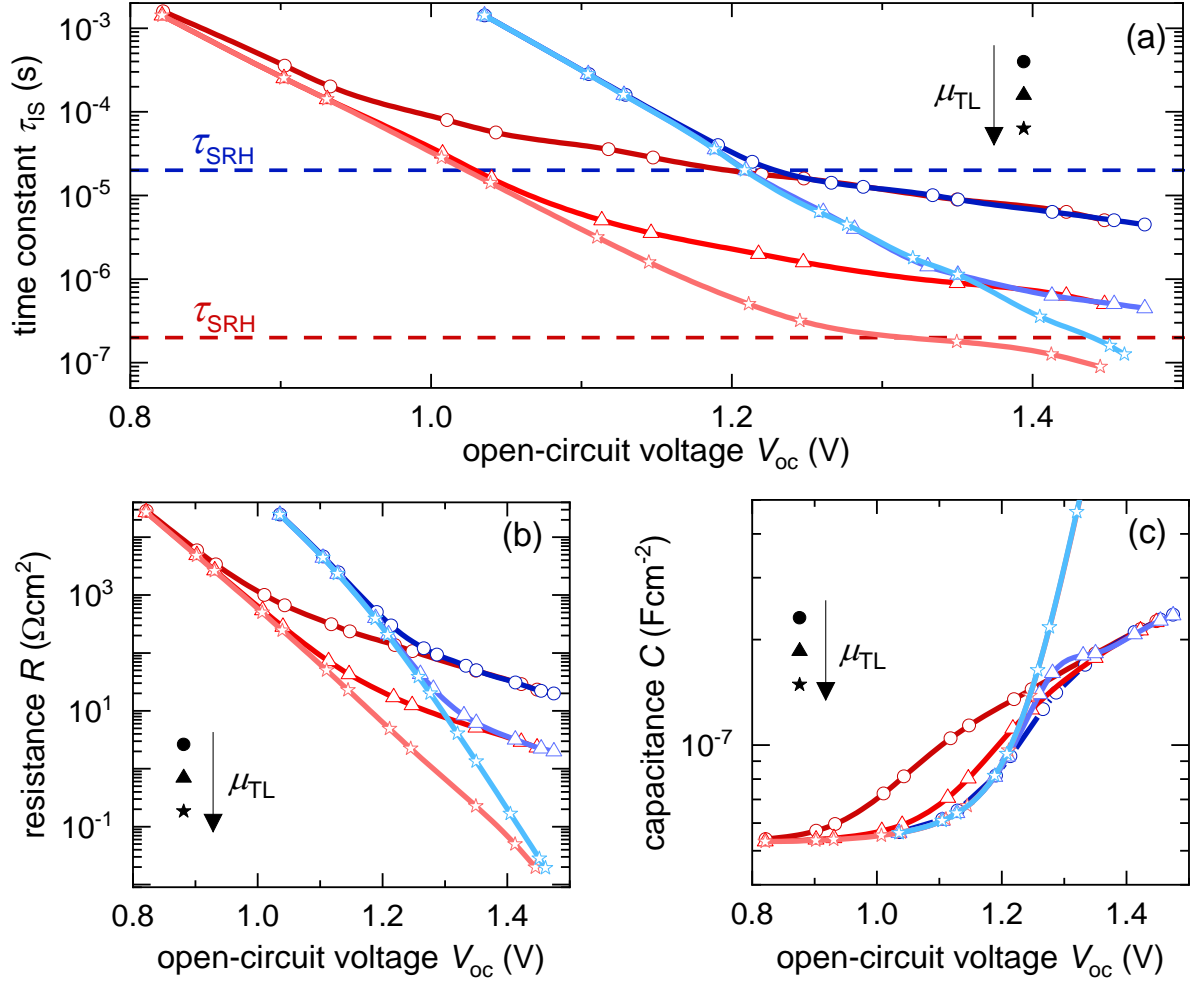

**Figure 8** Calculated (a) high frequency time constants and corresponding (b) resistances and (c) capacitances from IS simulations at open-circuit for different SRH lifetimes (dashed blue and red lines in (a)) from drift-diffusion simulations of a PSC, where the HTL mobility was set equal to the perovskite layer while the ETL mobility was varied as shown in the legend (stars correspond to the ETL mobility being equal to that of the perovskite layer). Reduced darkness of blue and red lines also correspond to increasing mobility of the ETL layer for the respective SRH lifetime. Simulation parameters are shown in table S4 in the SI. The exchange resistance $R_{exc}$ dominates the resistance and time constant at high open-circuit voltages for lower mobilities of the ETL.

*Variable ETL and HTL mobility*

We finally simulate the scenario of variable mobilities of both the ETL and HTL, shown in figure S12 in the SI. The time constants, resistances and capacitances show an identical evolution as a function of mobility and open-circuit voltage to the case of a single transport layer with variable mobility (figure 8). We conclude that the salient features of the IS response of a full PSC with variable mobilities of the ETL and HTL layers is captured well by the model. However, the contribution of each transport layer to the observed resistance cannot be known from such an analysis. Therefore, measurements on PSCs with the same ETL but different HTLs or vice versa will allow determination of the transport layer whose properties are expressed in the IS measurement.

## 5. Discussion

Having established the validity of the model in the previous sections, we now proceed to interpret the experimental IS data in figure 1. We first analyse the three data points that do not show a plateau in the time constant – three *SAM-C$_{60}$/BCP* devices (samples 6, 7 and 8 in figure 1). The corresponding slope factors of the resistances and capacitances are shown in figure S13 in the SI. The slope factor of the resistance shows a value between 1-2 at lower open-circuit voltages that transitions to lower values between 0.2-1 at higher open-circuit voltages. This behaviour is similar to the transition between first order to second order recombination in figures 6(c) and 8(b), though it is unclear why the values go below 1. The slope factor of the capacitance also decreases continuously from values above 2 to values below 1, similar to a transition from the geometric capacitance to the chemical capacitance at high open-circuit voltages. We conclude that the recombination resistance $R_{\text{rec}}$ dominates the measured resistances for devices with the *SAM-C$_{60}$/BCP* combination of transport layers, which therefore does not contain any information regarding the transport layer properties.

The dominance of $R_{\text{rec}}$ in unison with the absence of a plateau region raises the question – under which conditions can the plateau region corresponding to the SRH lifetime be clearly observed? We therefore calculate the width $\Delta V_{\text{sat}}$ of the plateau region to identify the parameters that it depends on. From the model, for $\tau_{\text{exc}} \ll \tau_{\text{eff}}$ (figure 4(a)), we know that $\Delta V_{\text{sat}}$ is determined by the intercept of the time constant $R_{\text{rec}}C_g$ with $\tau_{\text{SRH}}$ on the left side and $\tau_{\text{rad}}$ with $\tau_{\text{SRH}}$ on the right side, leading to (see section A6 in the SI for derivation)

$$\Delta V_{\text{sat}} = \frac{k_B T}{q} \ln\left(\frac{C_{\mu,0}}{C_g} \frac{\tau_{\text{rad},0}}{\tau_{\text{SRH}}}\right), \tag{41}$$

where $C_{\mu,0}$ and $\tau_{\text{rad},0}$ are the equilibrium chemical capacitance and radiative lifetime respectively. Equation 41 implies that $\Delta V_{\text{sat}} \propto d^2$, which means $\Delta V_{\text{sat}}$ can be increased by increasing the thickness of the perovskite layer (thereby reducing the geometric capacitance). This effect is confirmed in the simulations corresponding to a thin and thick perovskite layer shown in figure 9, where the device corresponding to a thick perovskite layer shows an 'S'-shaped evolution of the time constant with a clear plateau at $\tau_{\text{SRH}}$, for large enough mobilities of the transport layer. Thus, moving from thin film devices to devices with a thick perovskite layer (lateral devices or single crystals) opens up an interesting possibility to clearly observe the SRH lifetime.

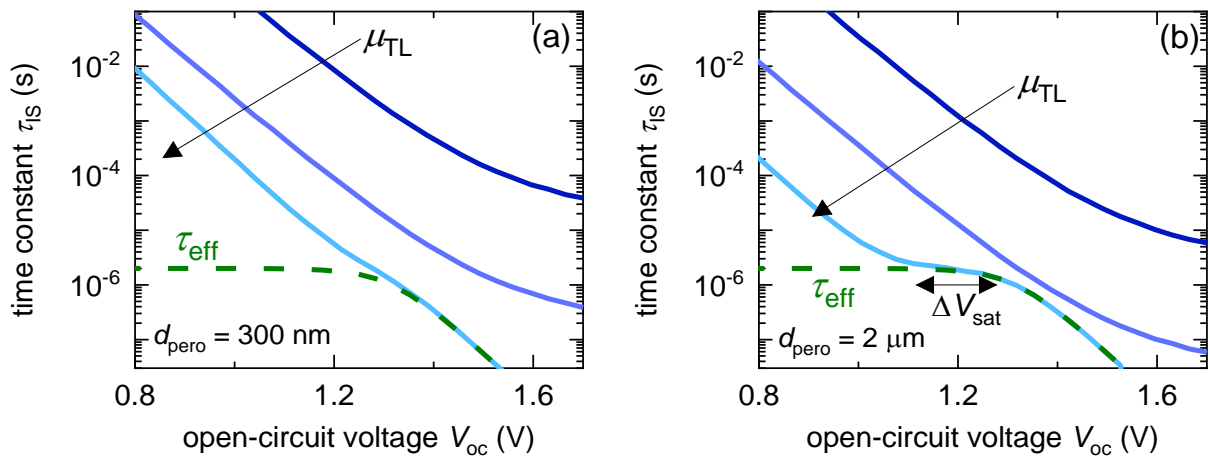

**Figure 9** Simulated time constants from the model for different mobilities of the transport layer for (a) a thin perovskite layer of thickness 300 nm and (b) a thick perovskite layer of thickness 2 µm. For lower mobilities of the transport layer, the time constants are dominated by the transport layer resistance $R_{\text{exc}}$, while for higher mobilities, the bulk recombination lifetime is expressed. The width of the plateau of the time constant that saturates at the SRH lifetime $\tau_{\text{SRH}}$ is increased for the device with a thick perovskite layer (equation 41). Simulation parameters

are shown in table S3 in the SI.

We now analyse the data points that show a plateau in the time constant in figure 1. These correspond to samples 1-5, that include the *PTAA-PCBM/BCP*, *PTAA-PCBM:CMC:ICBA/BCP* and the *PTAA-C$_{60}$/BCP* combinations of transport layers. The corresponding slope factors of the resistances and capacitances are shown in figure S13 in the SI. The *PTAA-PCBM/BCP* devices (samples 1 and 2) show a transition of the resistance slope factor from a value ~2 at low open-circuit voltages to much larger values that correspond to a saturation of the resistance at large open-circuit voltages. This behaviour is identical to the evolution of $R_{exc}$, shown in figure 8(b) and figure S10 in the SI. If the plateau region in the time constant indeed corresponds to a real SRH lifetime, then the corresponding resistance would have shown an exponential trend throughout the entire range of open-circuit voltage, without any plateau, as shown in figure 7(b). A similar trend is observed for the resistance slope factors of samples 3-5 corresponding to the *PTAA-PCBM:CMC:ICBA/BCP* and *PTAA-C$_{60}$/BCP* combination of transport layers. However, for the case $\tau_{exc} \cong \tau_{eff}$ shown in figure S6 in the SI, we observe that while $R_{exc}$ dominates the plateau region of the time constants at large open-circuit voltages, $R_{rec}$ dominates the exponential rise in the time constant at lower open-circuit voltages. Since we have no information regarding the value of $\tau_{SRH}$, we conclude that the measured resistance at high frequencies for samples 1-5 are dominated by the exchange resistance $R_{exc}$ only in the plateau region at large open-circuit voltages. This allows calculation of the exchange velocity $S$ using equations 15 and 6 as

$$S_{exc} = \frac{2k_B T}{q^2 n_{int} R_{exc}} \exp\left(-\frac{qV_{oc}}{2k_B T}\right). \tag{42}$$

Since samples 1-5 all correspond to the same HTL, we speculate that the observed $R_{exc}$ values originate from the *PTAA* layer. From the bandgaps of the perovskite layers for samples 1-5 (table S1 in the SI), we calculate (assuming a valence band density of states of $N_v = 10^{18}$ cm$^{-3}$) an intrinsic concentration of $n_i = p_i \cong 4.336 \times 10^4$ cm$^{-3}$ for the perovskite layer in samples 1,2,4 and 5 and $n_i = p_i = \cong 4.314 \times 10^3$ cm$^{-3}$ in sample 3. We further assume a conduction band offset $\Phi_b$ of 0.08 eV at the *perovskite/PTAA* interface,[42] which allows calculation of the interface concentration of holes (using equation 6) as $p_{int} \cong 9.4055 \times 10^5$ cm$^{-3}$ for samples 1,2,4 and 5 and $p_{int} \cong 9.3576 \times 10^4$ cm$^{-3}$ for sample 3. The exchange lifetime $\tau_{exc}$ is then calculated using equation 30 assuming a thickness of $d$=300 nm for the perovskite layer. If the electric field $F_{TL}$ in the transport layers was known, we could then also estimate the mobilities in the transport layers from $S_{exc}$ using equation 7. However, the electric field itself is very difficult to measure and estimates of the built-in electrostatic potential difference are generally quite crude in situations where the Mott-Schottky analysis fails (as is the case for perovskites).[18] We assume that the built-in electrostatic potential difference in the transport layers $V_{bi,TL}$ is comparable to the best open-circuit voltage values obtained for this band gap, since charge collection losses are maximised when the applied voltage is larger than the built-in electrostatic voltage.[36] Based on the 1 sun open-circuit voltage values and bandgaps for samples 1-5 in table S1 in the SI, we thus assume that $V_{bi,TL}$ of samples 1-3 is approximated by the respective open-circuit voltages. Therefore, for samples 1-3 at 1 sun open-circuit voltage conditions, we assume a negligible electric field in the transport layers, which allows approximating the exponential in equation 7 with a first-order Taylor expansion, yielding

$$S_{exc}(V_{oc}(1\text{ sun})) \cong \frac{\mu_{TL} F_{TL}}{1-\left(1-\frac{qF_{TL}d_{TL}}{k_B T}\right)} = \left(\frac{k_B T}{q}\right)\frac{\mu_{TL}}{d_{TL}}. \tag{43}$$

Equation 43 thus allows calculation of the mobility of the transport layer from the $S_{exc}$ value measured at 1 sun open-circuit voltage only for samples 1-3. Figure 10 shows the calculated $S$ values and the corresponding $\tau_{exc}$ and $\mu_{TL}$ values, assuming a thickness $d_{TL} = 20$ nm for the transport layer. The $S_{exc}$ values for samples 1-3 (*PTAA-PCBM/BCP* and *PTAA-PCBM:CMC:ICBA/BCP*) lie between 1-40 cm/s with corresponding $\tau_{exc}$ values approximately between 0.8-12 µs, while samples 4-5 (*PTAA-C$_{60}$/BCP*) show $S_{exc}$ values between 20-100 cm/s with corresponding $\tau_{exc}$ values between 0.3-1 µs. The calculated transport layer mobilities for samples 1-3 lie between $10^{-4}$-$3\times10^{-3}$ cm$^2$V$^{-1}$s$^{-1}$, similar to those reported in refs.[15-17] for PTAA and PCBM layers.

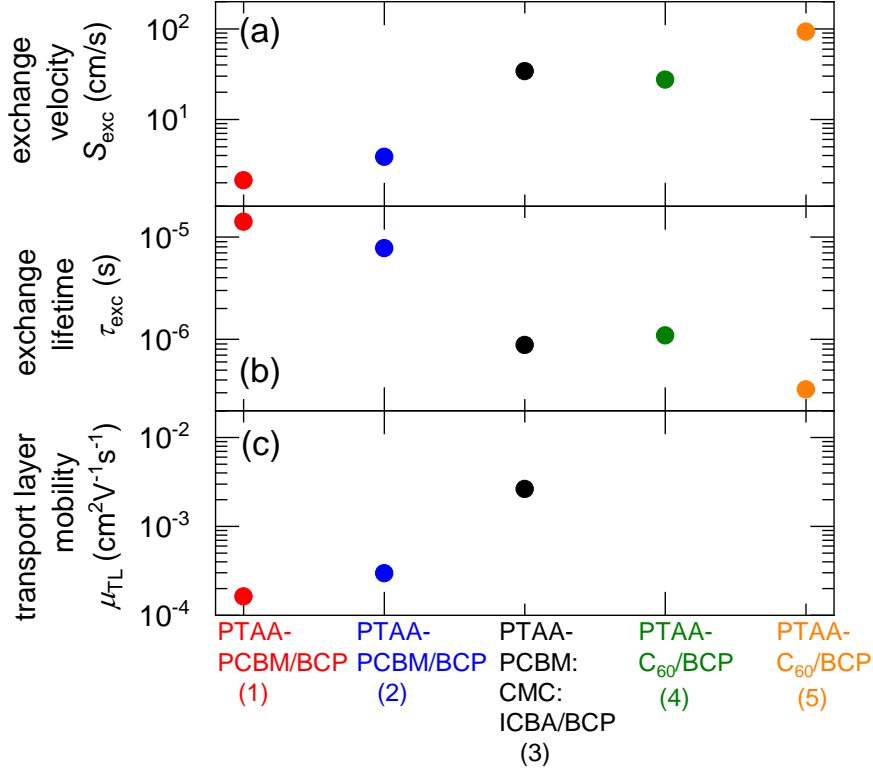

**Figure 10** Calculated charge carrier (a) exchange velocity $S_{exc}$, (b) exchange lifetime $\tau_{exc}$ and transport layer mobilities for samples 1-5 from the experimental IS data measured at 1 sun open-circuit conditions in figure 1, which correspond to the *PTAA-PCBM/BCP*, *PTAA-PCBM:CMC:ICBA/BCP* and *PTAA-C$_{60}$/BCP* combination of transport layers. A transport layer thickness $d_{TL} = 20$ nm was assumed to calculate the transport layer mobilities in (c).

## 6. Conclusions

In any time domain or frequency domain measurement of a solar cell, recombination and transport mechanisms are superimposed. In the case of perovskite solar cells, these mechanisms include not only transport within the perovskite layer but also charge extraction from the perovskite to the transport layer, followed by transport within the transport layer and subsequent collection by the electrode. Due to the large diffusion length of electronic carriers in the perovskite layer, the charge extraction and transport properties of the transport layers are significant factors that determine the charge collection efficiency of the perovskite solar cell. While models that account for charge transport within the absorber layer exist in the literature, there is a lack of knowledge in the photovoltaic community regarding the physical mechanisms that govern charge extraction and collection, and their coupling with recombination in the perovskite layer.

To solve this problem, we have developed a model for perovskite solar cells (and any multilayer solar cell in general) that describes the charge carrier exchange between the perovskite absorber and collecting electrodes through the transport layers ie. the conversion between the internal and external voltage. The model predicts the influence of a charge carrier exchange lifetime that determines the rate at which charge carriers are injected into or extracted from the bulk volume of the absorber layer through the transport layers. This lifetime is associated to a charge carrier exchange resistance that models the potential drop across the transport layers. For a variety of small-perturbation measurements in the frequency domain (IMVS, IMPS and IS), we find that the additional information required to decouple recombination and charge carrier exchange is embedded in the voltage dependence of the time constants, which can be easily measured experimentally. With the help of drift-diffusion simulations, we explain the coupling between the different physical mechanisms that generate the expressed and un-expressed time constants, providing both analytical solutions and an equivalent circuit for the interpretation of experimental data.

In the case of impedance spectroscopy, the model predicts that in cases where charge exchange is faster than the effective recombination lifetime, the typical 'S'-shaped evolution of the measured time constants is observed. The voltage dependence of the time constants in such a scenario is dominated by the coupling between the recombination resistance and geometric capacitance at low open-circuit voltages, followed by a plateau corresponding to the SRH lifetime and subsequent exponentially voltage-dependent radiative lifetime at high open-circuit voltages. For cases where the charge carrier exchange is slower than the effective recombination lifetime, the time constant is dominated by the coupling between the charge carrier exchange resistance and the geometric capacitance, showing a characteristic exponential evolution at low open-circuit voltages and a plateau at higher open-circuit voltages. This plateau can be mistaken as a SRH recombination lifetime if the corresponding resistances and capacitances are not analysed.

These insights were used to analyse impedance spectroscopy data of *p-i-n* perovskite solar cells with different transport layers. We find that the devices with the hole transport layer *PTAA* showed the characteristic behaviour and evolution of the charge carrier exchange resistance versus open-circuit voltage, irrespective of the type of electron transport layer used. This allowed determination of charge carrier exchange velocities between 1-100 cm/s at 1 sun open-circuit conditions, which corresponds to transport layer mobilities between $10^{-4}$-$3\times10^{-3}$ cm$^2$V$^{-1}$s$^{-1}$ for the *PTAA*-based PSCs.

**Author contributions**
S.R. carried out all the experiments, simulations and wrote the first draft and the final version of the manuscript. Z.L. and Y.W. fabricated the samples. U.R. developed the concept of the matrix model and contributed to the writing of the manuscript. T.K. supervised the project and contributed to reviewing and editing the manuscript.

**Conflicts of interest**
There are no conflicts of interest to declare.

**Data availability**
Drift-diffusion simulations were performed using SETFOS developed by Fluxim AG and matrix model simulations were performed using MATLAB. Analytical solutions of the matrices were obtained using Maplesoft.

**Acknowledgements**
S.R. acknowledges the German Research Foundation (DFG) for the Walter-Benjamin fellowship – project number 462572437. U.R. and T.K. acknowledge funding from the Helmholtz association via the project PEROSEED. Open access publication funded by the German Research Foundation (DFG) – 491111487.